\documentclass[
final
, nomarks
]{dmtcs-episciences}

\usepackage[utf8]{inputenc}
\usepackage{subfigure}

\usepackage[round]{natbib}
\usepackage{amsmath}

\usepackage{amssymb}
\usepackage{amsmath}
\usepackage{amsthm}
\usepackage{clrscode3e}
\newcommand{\N}{\naturals}
\newcommand{\Z}{\integers}
\newcommand{\Q}{\mathbb{Q}}
\newcommand{\structure}[1]{$\{#1\}$}
\newcommand{\structureb}[2]{$\{#1\}^{#2}$}

\newcommand{\sdp}{\textsc{set-red }}
\newcommand{\sdpQ}{\textsc{set-red}}
\newcommand{\mdp}{\textsc{multiset-red }}
\newcommand{\mdpQ}{\textsc{multiset-red}}
\newcommand{\npf}{\textsc{$\N$-poly-red }}
\newcommand{\npfQ}{\textsc{$\N$-poly-red}}

\newcommand{\multiset}[1]{\{\!\{ #1 \}\!\}}
\newcommand{\singleset}[1]{\{ #1 \}}
\newcommand{\mink}{\oplus}

\newtheorem{definition}{Definition}
\newtheorem{exam}{Example}
\newtheorem{theo}{Theorem} 
\newtheorem{quest}{Question}

\newcommand{\Init}{
\begin{codebox}
\Procname{$\proc{InitialSolution}(M,n)$}     
\li $m \gets  \proc{Round}(\attrib{M}{length}/2)$ 
\li $M \gets \proc{Sort}(M)$ 
\li $M \gets M[1 \twodots m]$
\li $M \gets \proc{RandomSample}(M,n)$
\li  \Return $M$
 \end{codebox}
}

\newcommand{\Score}{
\begin{codebox}
\Procname{$\proc{Score}(M,S)$}     
\li \Comment invariant: $S[1]=0$, $S \subseteq M$ and $\attrib{S}{length}$ divides $\attrib{M}{length}$
 \li $\id{col} \gets \attrib{S}{length}$ 
\li $\id{row} \gets \attrib{M}{length}/col$ 
 \Comment Let $mat$ be an $row \times col$ matrix whose entries are set to $0$
 \li $\id{r} \gets M\setminus S$
 \li \Comment first row of $mat$ gets $S$
\li $score \gets col$
 \li 	\For $\id{i} \gets 2$ \To $row$  		\Do 
	 \li 	 $w \gets$ \proc{Min($r$)}	
	 \li 	  $r \gets r \setminus \{w\}$ 
	\li $score \gets score +1$
  \li \Comment $mat[row,1] \gets w$
 \li 	\For $\id{j} \gets 2$ \To $col$ 		\Do 
 \li $\id{c} \gets w+S[j]$
\li \If $c \in r$ 
	 \li \Then 
	  $\id{r} \gets  r \setminus \{c\}$ 
	\li $score \gets score +1$
	 \li \Comment $mat[row,col] \gets c$
	 \li \Else   \Return $score$ \End                    
      \End
\End
\li   \Return \id{score}	
\end{codebox}
}

\newcommand{\NSearch}{
\begin{codebox}
\Procname{$\proc{NeighborSearch}(M,S)$}     
 \li \Comment invariant: $S[1]=0$ and $\attrib{S}{length}$ divides $\attrib{M}{length}$
 \li $initial\_score \gets \proc{Score}(M,S)$ 
 \li 	  $alternatives \gets \proc{DeleteDuplicates}( M \setminus S)$ 
 \li 	\For $\id{i} \gets 2$ \To $\attrib{S}{length}$  		\Do  
	 \li 	\For $\id{j} \gets 1$ \To $ \attrib{alternatives}{length}$ 		\Do 
		 \li $temp \gets S[i]$
		  \li $S[i] \gets alternatives[j]$
		 \li $new\_score \gets \proc{Score}(M,S)$
		\li \If $new\_score > initial\_score$ 
	 \li \Then  \Return $(new\_score,S)$ 
	 \li \Else $S[i] \gets temp$ \End 
       \End
\End
\li   \Return $(initial\_score,S)$
\end{codebox}
}

\newcommand{\LocOpt}{
\begin{codebox}
\Procname{$\proc{FindLocalOpt}(M,S)$}     
\li \Comment invariant: $S[1]=0$
 \li $\id{n} \gets \attrib{M}{length}$ 
 \li $current\_score \gets \proc{Score}(M,S)$ 
 \li 	\While \const{true}  		\Do 
		 \li 	  $(score,S) \gets \proc{NeighborSearch}(M,S)$	 
		 \li \If $score \isequal n$ 
	 \li \Then \Return $(\const{true},S)$ \End      
	 \li \If $score \isequal current\_score$ 
	 \li \Then \Return $(\const{false},S)$ \End      
	 \li $current\_score \gets score$
\End
\end{codebox}
}

\newcommand{\Iterated}{
\begin{codebox}
\Procname{$\proc{IteratedSearch}(M,m,iterations)$}     
\li \Comment invariant: $m$ divides $\attrib{M}{length}$
\li $current\_solution \gets \proc{InitialSolution}(M,m)$ 
 \li 	\For $i \gets 1 $ \To $iterations$  		\Do  
  \li 	  $(found,S) \gets \proc{FindLocalOpt}(M,current\_solution)$	 
  \li \If $found$ 
		 \li \Then \Return $S$ \End    
\li $current\_solution \gets  \proc{NewInitialSolution}(M,current\_solution)$
  \li \Comment note that $current\_solution$  contains $0$
 \End
\li  \Return solution not found
 \end{codebox}
}

\newcommand{\Perturb}{
\begin{codebox}
\Procname{$\proc{NewInitialSolution}(M,S)$}     
\li \Comment invariant: $S[1]=0$, all the elements of $S$ are in $M$ and $\attrib{S}{length}$ divides $\attrib{M}{length}$
 \li $\id{col} \gets \attrib{S}{length}$ 
\li $\id{row} \gets \attrib{M}{length}/col$ 
\li \Comment Let $mat$ be an $row \times col$ matrix whose entries are set to $0$
 \li $\id{R} \gets M\setminus S$ 
 \li \Comment first row of $mat$ gets $S$
\li $new\_set \gets S$
 \li 	\For $\id{i} \gets 2$ \To $row$  		\Do 
	 \li 	 $w \gets$ \proc{Min($R$)}	
	 \li 	  $R \gets R\setminus \{w\}$  
	\li $new\_set  \gets new\_set  \bigcup \{w\}$
  \li \Comment $mat[row,1] \gets w$
 \li 	\For $\id{j} \gets 2$ \To $col$ 		\Do 
 \li $\id{c} \gets w+S[j]$
\li \If $c \in R$ 
	 \li \Then 
	  $\id{r} \gets R\setminus \{c\}$ 
	   \li \Comment $mat[row,col] \gets c$
	 \li \Else   \Return $\proc{RandomSample}(new\_set,col)$ 
	  \li \Comment  $\proc{RandomSample}(new\_set,col)$ must contain $0$
	 \End                    
      \End
\End
\li   \Return $\proc{RandomSample}(new\_set,col)$ 	
\end{codebox}
}

\author{Luciano Margara}
\title[Decomposing
multisets according to the Minkowski sum]{A heuristic technique for decomposing
multisets of non-negative integers according to the Minkowski sum}
\affiliation{
 Department of Computer Science and Engineering,
 University of Bologna, Italy}
\received{2022-08-02}

\revised{2022-10-10}

\accepted{2022-10-25}

\begin{document}
\publicationdetails{24}{2022}{2}{11}{9877}

\maketitle
\begin{abstract}
 We study  the following problem.
Given a multiset $M$ of  non-negative integers, decide whether there exist
and, in the positive case, compute two non-trivial multisets whose Minkowski 
sum is equal to $M$.
The Minkowski sum of two multisets $A$ and $B$  is a multiset containing all possible sums of any element of $A$ and any element of $B$.
This problem was proved to be NP-complete when multisets are replaced by sets. This version of the problem is strictly related  to the factorization of boolean polynomials that turns out to be NP-complete as well. When multisets are considered, the problem is equivalent to the factorization of polynomials with non-negative integer coefficients.
The computational complexity of both these problems is still  unknown. 

The main contribution of this paper is a heuristic technique for decomposing multisets of non-negative integers.
Experimental results show that our heuristic decomposes multisets of hundreds of elements within seconds, independently of the magnitude of numbers belonging to the multisets.
Our heuristic can also be used  for factoring polynomials in $\N[x]$.  We show that, when the degree of the polynomials gets larger, our technique is much faster than the state-of-the-art algorithms implemented in 
commercial software like Mathematica and  MatLab.

\end{abstract}

%DMTCS is an open access scientific is implemented by the
%\emph{episcience} platform, see \cite{berthaud:hal-01002815} for an
%overview of the strategy. It combines high scientific and editorial
%quality with an open access policy. It is priceless, neither authors
%nor readers pay money for the access. Access is granted by giving
%episcience an irrevocable license to publish the articles, the
%copyright remains with the authors. The platform itself is run by
%French government services that do their best to warrant continuous
%access and a high quality of service.
%
%This document describes the use of the \texttt{dmtcs-episcience.cls}
%document class. It should be used
%\begin{center}
%  \emph{\textbf{for all DMTCS publications}}.
%\end{center}
%
%If you are still preparing a document for our previous \texttt{OJS}
%platform, please add \texttt{ojs} to the classes options, see
%Section~\ref{sec:options}.

%=======================================================================
%=======================================================================
%=======================================================================
%=======================================================================
\section{Introduction}
\label{sec1}
%=======================================================================
%=======================================================================
%=======================================================================
%=======================================================================

The idea of decomposing a mathematical object  into the sum (product, or other operations) of  smaller ones
is definitely not new. 
A huge literature has been devoted to the factorization of numbers, polynomials, matrices, graphs and 
many other mathematical objects, including sets and multisets. 
The basic idea behind factorization is decomposing a complex object into smaller and easier to analyze pieces.
Properties satisfied by each piece might shed some light on the properties satisfied by the entire object.
As an example, from irreducible factors of a polynomial, we can recover valuable information about its roots.
In this paper, we study the decomposition of multisets of non-negative integers according to the 
Minkowski sum. 
Multisets are an extension of 
the notion of sets where, basically, multiple copies of the same element are allowed.
The Minkowski sum is a binary operation that can be applied both to sets and multisets.
The Minkowski sum of two multisets $A$ and $B$  is a multiset containing all possible sums of any element of $A$ and any element of $B$.

%\medbreak %\noindent
Given a multiset $M$ of  non-negative integers,
the decomposition problem asks for computing two 
non-trivial multisets whose Minkowski sum is equal to $M$.
%\medbreak
Multisets theory have applications  in many fields~\cite{Singh2007}, e.g.,  in combinatorics~\cite{anderson2002combinatorics,stanley_2011,stanley_fomin_1999},
in the theory of relational databases~\cite{GrumbachM96,friz22,lamp00}, 
in multigraphs theory~\cite{DeVos2013,DUDEK2013785} and in computational geometry ~\cite{Ioa17}.
The problem of decomposing multisets of non-negative integers
is strictly related to the problem of factoring  univariate polynomials 
with non-negative coefficients (see Section \ref{sec2.5} for details).
Even if this problem 
arises in a very natural way in a number of different theoretical and practical contexts, it has not been thoroughly studied (see for example~\cite{Brunotte,fac19,Woestijne}) and its computational complexity is still unknown.
To our knowledge, no polynomial time algorithm  nor an NP-completeness proof exists.
When multisets are replaced by sets, the decomposition  problem was proved to be NP-complete~\cite{Kim05}. 
Other variants of the Minkowski sum decomposition problem have been studied. As an example, 
in~\cite{Gao01} the authors study the Minkowski decomposition of integral convex polytopes proving that 
the decisional version of this problem is again NP-complete.

%\medbreak%\noindent
The main contribution of this paper is a heuristic technique for decomposing multisets of non-negative integers which, in turn, can be applied to factoring polynomials with non-negative coefficients.

%\medbreak
The idea behind our algorithm is to transform the decomposition problem 
in an optimization problem by introducing a score function for
candidate solutions. A candidate solution is an approximation of a solution. 
The score function measures the quality of candidate solutions, i.e., the similarity to the actual solution (not necessarily unique). The score function reaches its maximum (whose value is known in advance) only at a solution for the problem.
Our algorithm  starts from a randomly generated candidate solution $s_0$ and iteratively improves it until it finds 
a local optimum candidate solution $s_k$ according to the score function. If $s_k$ 
reaches the maximum score the algorithms terminates, otherwise it starts over from another  initial candidate solution computed starting from $s_k$.
The maximum number of  iterations is bounded by a predetermined threshold.

%\medbreak
We extensively tested our algorithm over randomly generated instances of different size and 
structure. 
Experimental results (see Section \ref{sec2.7} and Tables in Appendix \ref{data} and \ref{data2}) show that after a small number of iterations our algorithm 
almost always finds a solution.

%\medbreak
As far as polynomial with non-negative coefficients factorization  is concerned,
no  efficient and specifically designed algorithms are known.
A possible natural strategy to solve this problem  might consist of factoring the polynomial 
in $\Z[x]$ (this can be done in polynomial time) and then
suitably grouping factors in $\Z[x]$ 
in order to get factors in $\N[x]$. 
Unfortunately, there exists no 
efficient algorithm to perform the grouping of factors whose number can be, in general,  exponentially large.
In our opinion, this is an interesting problem in itself.
Since decomposing multisets of non-negative integers 
is equivalent (under some conditions we will discuss in Section \ref{sec2.5}) to the problem of factoring
polynomials in $\N[x]$, 
the alternative strategy might also be used  for decomposing multisets.
In Section \ref{sec4.5} we make a comparison between our algorithm and the alternative strategy depicted above unrealistically assuming
that the grouping of factors can be computed for free. We used built-in functions 
provided in Wolfram Mathematica language for integer 
polynomials factorization (similar results have been found using MatLab). 

%\medbreak%\noindent
Experimental results clearly show (see Tables \ref{t13},\ref{t14} and \ref{t15} in Appendix \ref{data2}) that, when the degree of polynomials increases, our technique 
is much faster  than going through factoring. 
Reversing the line of reasoning, i.e., using multisets decomposition techniques for factoring polynomials in $\N[x]$,
our heuristics becomes a serious candidate to be the first effective method for factoring polynomials
with non-negative coefficients.

%\medbreak
The rest of this paper is organized as follows. In Section \ref{sec2} we 
give basic definitions and known results. In Section \ref{sec3} we describe our heuristics and we provide 
its pseudocode.
In Section \ref{sec2.7} we show experimental results. 
In Section \ref{sec4.5} we make a comparison between our algorithm and an alternative strategy for decomposing multisets based on integer polynomial factorization. 
Section \ref{sec5} contains conclusions and some ideas for further works. 
Appendices \ref{data} and \ref{data2} contain tables with experimental data.

%=======================================================================
%=======================================================================
%=======================================================================
%=======================================================================
\section{Definitions and Known Results}\label{sec2}
%=======================================================================
%=======================================================================
%=======================================================================
%=======================================================================

Let $\Z$ be the set of  integers and $\Z[x]$ be  the sets of univariate 
polynomials with coefficients in $\Z$.
Let $\N$ be the set of non-negative  integers and $\N[x]$ be  the sets of univariate 
polynomials with coefficients in $\N$.

Multisets are an extension of the concept of sets.
While a set can contain only one occurrence of any given element, 
a multiset may contain multiple occurrences of the same element.
To distinguish multisets from sets, we will represent multisets by using double braces.

As an example $M=\multiset{2, 2, 3, 3, 5, 5, 5, 5, 5, 6, 8, 8}$ is a multiset. 
Given a multiset $M$ we denote by $\mu(x,M)$ the number of occurrences (possibly $0$) 
of the element $x$ in $M$.
Sometimes we will represent a multiset $M$ as a set of pairs $(element,\mu(element,M)$.
With this notation, the above multiset   can be  written as 
$M=\singleset{(2,2),(3,2),(5,5),(6,1),(8,2)}$.
In what follows, we will consider sets and multisets of  numbers. This enable us to define a binary operation 
on them (denoted by the symbol $\mink$)  
sometimes called Minkowski sum. We will use the symbol $\mink$
both for sets and multisets sum inferring the type of operation from the type of
operands. 

 \begin{definition}[Minkowski Set Sum]\label{sum}
The Minkowski sum  of 
two sets $A$ and $B$ is a set defined as follows.
$$ A \mink B = \singleset{ a+b:\ a\in A \text{ and } b\in B}
$$
\end{definition}

\begin{exam}\label{ex:sum}
Example of set sum. Let $A=\singleset{0,1,3}$ and  $B=\singleset{2,5}$. Then
$A\mink B=\singleset{2,3,5,6,8}$. Since we are working with sets, the multiplicity of $5$
in $A\mink B$ is $1$ even if 
$5$ can be obtained both as $0+5$ and $3+2$.
\end{exam}

\begin{definition}[Minkowski Multiset Sum]\label{msum}
The Minkowski sum  of 
two multisets  $A$ and $B$ is a multiset given by
$$ A \mink B = \multiset{ a+b:\ a\in A \text{ and } b\in B}
$$
\end{definition}

\begin{exam}\label{ex:msum}
Examples of multiset sum. \\
Let $A=\multiset{0,1,3}$ and  $B=\multiset{2,5}$. Then
$A\mink B=\multiset{2,3,5,5,6,8}$.\\
Let $A=\multiset{0,1,3,3}$ and  $B=\multiset{2,2,5}$. Then%\\
$A\mink B=\multiset{2, 2, 3, 3, 5, 5, 5, 5, 5, 6, 8, 8}$.
\end{exam}
The identity element with respect to the set sum is $\singleset{0}$
and the identity element with respect to the multiset sum is
$\multiset{0}$.
A multiset $A$ is contained in a multiset $B$ ($A\subseteq B$) if and only if 
\begin{equation}\label{subseteq}
\forall x\in A: \ \  x\in B \text{ and } \mu(x,A) \leq \mu(x,B)
\end{equation}
We also define the multiset difference operation (denoted by the  $\setminus$ symbol) as follows.
\begin{equation}\label{difference}
A\setminus B =\singleset{(x,m_x):\  x\in A \text{ and } m_x =\max(\mu(x,A)-\mu(x,B),0)  }
\end{equation}

%\noindent
As an example, $\multiset{2, 2, 3, 3, 5, 6, 8, 8}\setminus \multiset{2, 3, 3,3, 5, 9}=\multiset{2, 6, 8, 8}$.
%\medbreak%\noindent
We now introduce the notion of reducible multisets (sets) of non-negative integers.
 
 \begin{definition}[Reducible multiset (set)]\label{red}
A multiset (set) $M$ of non-negative integers is reducible if and only if there exist
two multisets (sets) $A$ and $B$, both of them different from the identity element,
 such that 
 $M=A\mink B$.
 \end{definition}
A multiset (set) $M$ of non-negative integers is irreducible (sometimes called prime) 
if and only if it is not reducible.
We are now ready to  state the following two problems.

 \begin{definition}[\sdpQ]\label{setdec}
Given a set $S$ of non-negative integers, decide whether $S$ is reducible or not.
\end{definition}

 \begin{definition}[\mdpQ]\label{msetdec}
Given a multiset $M$ of non-negative integers, decide whether $M$ is reducible or not.
\end{definition}
The following result was proved in \cite{Gao01}.
\begin{theo}\label{sdpNPcomplete}
\sdp is NP-complete.
\end{theo}

Unlike \sdpQ, the computational complexity of \mdp is, to our knowledge, still unknown.
This leads us to state the following open question.

\begin{quest}\label{mdpNPcomplete}
Is \mdp  NP-complete ?
\end{quest}

Even if we have defined \sdp and \mdp in their  decisional version, 
in the rest of this paper we will refer to them (with a little abuse of notation)
as constructive problems, i.e,  the problem of effectively computing
 two multisets (sets) whose Minkowski sum is equal to the multiset (set) received as input.
 
 In the next example we show that the irreducible factorization 
 of non-negative integer multisets  is not unique. This 
 makes the problem of factoring multisets even harder, if possible.

\begin{exam}\label{notunique}
Let $M=\multiset{0 ,1 ,2 ,3,4,5}$. Then 
\begin{eqnarray*}
M &=& \multiset{0 ,1} \mink\multiset{0 ,2 ,4}\\
	&=&  \multiset{0 ,3} \mink\multiset{0 ,1 ,2}.
\end{eqnarray*}
Multisets  $\multiset{0 ,1}, \multiset{0 ,2 ,4}, \multiset{0 ,3}$ and  $\multiset{0 ,1 ,2}$ 
are irreducible.
\end{exam}

%=======================================================================
%=======================================================================
%=======================================================================
%=======================================================================
\subsection{Multisets decomposition and polynomials factorization}\label{sec2.5}
%=======================================================================
%=======================================================================
%=======================================================================
%=======================================================================

One of the most studied problem in computer algebra is the problem of factoring 
polynomials.
A huge literature has been devoted to the factorization of polynomials (without claim of exhaustiveness see    \cite{VANHOEIJ2002167,Lenstra82factoringpolynomials,DBLP:conf/latin/Kaltofen92}).
The first polynomial factorization algorithm was published by Theodor Von Schubert in 1793 \cite{Schubert}.
Since then, dozens of papers on the computational complexity of polynomial factorization have been published.
In 1982, Arjen K. Lenstra, Hendric W. Lenstra, and L\'aszl\'o Lov\'asz \cite{Lenstra82factoringpolynomials} published the first polynomial time algorithm for factoring 
polynomials over $\Q$ and then over $\Z$.

The problem of factoring polynomials over a ring can be, in a sense,  labeled as ``well studied" and ``efficiently solved". The same cannot be said when rings are replaced by semirings (e.g.    the natural numbers).
Unlike the case of factoring polynomials over rings, the problem of factoring polynomials over semirings 
has received far less attention, there are far fewer known results 
and many interesting unanswered questions. One of them is the following.

\begin{quest}[\npfQ]\label{npf}
Given a polynomial $p(x)\in \N[x]$, decide whether $p(x)$ is reducible in $\N[x]$.
\end{quest}

As far as we know, 
for the \npf problem, there are neither polynomial algorithms to solve it nor proofs of NP-completeness.
\npf problem is strictly related to the \mdp problem.
%\medbreak

To any given polynomial $p(x)\in \N[x]$ it is possible to associate a multiset 
 as follows.
Let $p=a_0+a_1 x+a_2 x^2+\cdots + a_n x^n$ be any element of $\N[x]$.
We define the multiset 
\begin{equation}\label{mul}
Multiset(p)=\multiset{\overbrace{0,\dots,0}^{a_0},\dots,\overbrace{i,\dots,i}^{a_i},\dots,
\overbrace{n,\dots,n}^{a_n}}
\end{equation}

On the other hand, we can associate to any multiset 
$$M=\multiset{\overbrace{n_1,\dots,n_1}^{m_1},\overbrace{n_2,\dots,n_2}^{m_2},\dots,
\overbrace{n_d,\dots,n_d}^{m_d}}$$ the polynomial
\begin{equation}\label{pol}
Polynomial(M)=m_1 x^{n_1}+m_2 x^{n_2}+\cdots + m_d x^{n_d} 
\end{equation}

%\noindent
It is not difficult  to verify  that\\ 
- $Polynomial(Multiset(p))=p$
 and 
 $Multiset(Polynomial(M))=M$\\
- $Multiset(p\,q)=Multiset(p)\mink Multiset(q)$ and \\
- $Polynomial(A\mink B)=Polynomial(A)\,Polynomial(B)$

%\medbreak
%\noindent
As a consequence of these properties we have that\\
- $M$ is an irreducible multiset of non-negative integers if and only if \\ $Polynomial(M)$ is an irreducible 
polynomial over $\N[x]$ and\\
- $p$ is an irreducible 
polynomial over $\N[x]$ if and only if $Multiset(p)$ is an irreducible multiset of non-negative integers.

Unfortunately, in the general case, 
the size of $Multiset(p)$ may be exponentially larger than the size of $p$. This prevents us from  readily translating  computational complexity 
results for \mdp into equivalent results for \npf and viceversa. 

Taking advantage of  Example \ref{notunique}
we show that the irreducible factorization 
 of polynomials in $\N[x]$  is not unique. 

\begin{exam}\label{notuniqueP}
Let $p(x)=1 + x + x^2 + x^3 + x^4 + x^5$. 
The complete factorization of $p(x)$ in $\Z[x]$ is $p(x)=(1 + x) (1 - x + x^2) (1 + x + x^2)$.
Since $(1 + x) (1 - x + x^2)\in \N[x]$ and $(1 - x + x^2) (1 + x + x^2)\in \N[x]$,  then we have two distinct factorizations of $p(x)$ in $\N[x]$.
\begin{eqnarray*}
p(x) &=& (1 + x) (1 + x^2 + x^4)\\
	&=& (1 + x^3) (1 + x + x^2)
\end{eqnarray*}
\end{exam}

%=======================================================================
%=======================================================================
%=======================================================================
%=======================================================================
\section{The Heuristics}\label{sec3}
%=======================================================================
%=======================================================================
%=======================================================================
%=======================================================================

 In this section we provide a complete description of our heuristics by using pseudocode
  (for details see pages from $20$ to  $22$ in \cite{cormen}).
 
 %\medbreak
Given a multiset $M$  of $n$ non-negative integer numbers, a candidate  solution  for 
$M$ is any multiset $A$ ($A \neq \multiset{0}$) of cardinality $m$ such that  
$A\subseteq M$ and $m$ divides $n$.
A candidate  solution $A$ for $M$ is also a solution for $M$ if and only if 
there exists another  candidate  solution $B$  ($B \neq \multiset{0}$)  for 
$M$ such that $M=A\mink B$.
Given a candidate solution $A$ for $M$, deciding whether $A$ is also a solution for $M$
 can be done in polynomial time.
Given a  solution $A$ for $M$, computing $B$ such that $M=A\mink B$
 can be done in polynomial time.

 %\medbreak%\noindent
 Our heuristics starts from an initial candidate solution of a given cardinality 
and iteratively improves it (according to a given 
score function) until it finds a solution.
 %\medbreak
The cardinality $m$ of the initial candidate solution is unknown in advance but
must divide the cardinality of $M$.
For computing an actual decomposition of a multiset $M$ of cardinality $n$ we have to run
our algorithm on all possible factors $f$  of $n$ with $f\leq \sqrt{n}$.
We are aware that this  leads to an overhead of computation,
but luckily,  the number of factors of any positive  integer $n$ (not exceeding  $\sqrt{n}$) is 
very small if compared to $n$.
For every positive integer $n$, with $100\leq n\leq 100.000$,
we computed its number of factors  divided by $n$. 
It turns out that the average of these ratios is $0.00025$ and the maximum is $0.058$ 
(higher values are obtained for small numbers). 
For these reasons, in what follows, we will assume that the target cardinality of solutions is known.

%\medbreak
%\noindent
We now give the pseudocode of each function used in our heuristics
and  a short explanation on how it works.

\Init
%\medbreak%\noindent
\proc{InitialSolution} 
 takes as input a multiset $M$ and a non-negative integer $n$ 
that divides the cardinality of $M$ and returns a candidate solution of cardinality $n$.

\Score
%\medbreak%\noindent
\proc{Score} takes as input a multiset $M$ and a candidate solution $S$ for $M$ and
returns a positive integer 
measuring the quality of $S$. $\proc{Score($M,S$)}$ ranges from length of $S$ (lowest quality) to
 length of $M$ (highest quality). If $\proc{Score($M,S$)}=$ length of $M$ then $S$ is a solution for $M$.

To better understand how \proc{Score} works, we describe its behavior on the following example. 
Let $A=\multiset{0, 1, 3, 3}$, $B=\multiset{0, 2, 2, 6}$, and  $$M=A\mink B = 
\multiset{0, 1, 2, 2, 3, 3, 3, 3, 5, 5, 5, 5, 6, 7, 9, 9}$$

%\medbreak
%\noindent
Assume now to run \proc{Score($M,B$)}. Since $B$ is a solution for $M$, 
 \proc{Score($M,B$)} returns $16$, i.e., the length of $M$.
The matrix $mat$ described (but not computed)   at lines 6,8,14 and 20 would be
$$
mat= \left[
\begin{matrix}
0&2&2&6\\
1&3&3&7\\
3&5&5&9\\
3&5&5&9
\end{matrix}
\right]
$$
and the elements of $mat$ would give exactly the multiset $M$.

%\medbreak
%\noindent
Assume now to run \proc{Score($M,C$)}. Where $C=\multiset{0, 1, 2, 6}$ is a candidate solution but not a solution.
 \proc{Score($M,C$)} returns $6$.
The matrix $mat$ would now have  the form
$$
mat= \left[
\begin{matrix}
0&1&2&6\\
2&3&0&0\\
0&0&0&0\\% !TEX spellcheck = % !TEX spellcheck = 
0&0&0&0
\end{matrix}
\right]
$$
The element at row 2 and column 3 ($2+2=4$) in $mat$ cannot be found in $M$ (note that we have already removed 
$0,1,2,6,2$ and $3$ from $M$) and then 
 \proc{Score($M,C$)} stops at line 21 returning $6$, i.e.,  the number of elements correctly placed in $mat$ 
 until that moment.

%\medbreak
%\noindent
Last case. Assume  to run \proc{Score($M,C$)}. Where $C=\multiset{0, 2, 2, 5}$ is again a candidate solution but not a solution.
 \proc{Score($M,C$)} returns $11$.
The matrix $mat$ would have now the form
$$
mat= \left[
\begin{matrix}
0&2&2&5\\
1&3&3&6\\
3&5&5&0\\
0&0&0&0
\end{matrix}
\right]
$$
The element at row 3 and column 4 ($3+5=8$) in $mat$ cannot be found in $M$ and then 
 \proc{Score($M,C$)} stops at line 21 returning $11$, i.e., the number of elements correctly placed in $mat$ 
 until that moment.

\NSearch
%\medbreak%\noindent
\proc{NeighborSearch} takes as input a multiset $M$ and a candidate solution $S$ for $M$ and
 returns a candidate solution $N$ in the neighborhood of $S$ 
such that \proc{Score($M,N$)} $>$  \proc{Score($M,S$)}, if any.  Returns 
$S$, otherwise.

Given a multiset $M$ and a candidate solution $S$ for $M$, a neighbor
of $S$ is any candidate solution for $M$ differing from $S$ for exactly 1 element.
To speed up the process,  \proc{NeighborSearch} returns (line 11) the first improved candidate solution found.

\LocOpt
%\medbreak%\noindent
\proc{FindLocalOpt}  takes as input a multiset $M$ and a candidate solution $S$ for $M$ and
 returns a candidate solution $N$ with the property of being the best candidate solution in its neighbor, i.e.,
a local optimum. 
To accomplish this task, \proc{FindLocalOpt} keeps on calling \proc{NeighborSearch}
on improved solutions until no more improvement is found.
Note that the candidate solution  $N$ produced by  \proc{FindLocalOpt} is not guaranteed to be a solution.

\Iterated
%\medbreak%\noindent
\proc{IteratedSearch} takes as input a multiset $M$, an integer $m>1$ dividing the cardinality of $M$ and an upper bound on the number of iterations and returns a solution of cardinality $m$, if found. 
\proc{IteratedSearch} keeps on calling \proc{FindLocalOpt} with different initial candidate solutions
(computed by \proc{NewInitialSolution}) until a solution is found or the maximum number of iterations is exceeded.

\Perturb
%\medbreak%\noindent
\proc{NewInitialSolution} takes as input a multiset $M$ and a candidate solution $S$ for $M$ and returns a new initial candidate solution.
%\medbreak%\noindent
To better understand how \proc{NewInitialSolution} works, we show its behavior on an example. 
Let $A=\multiset{0, 1, 3, 3}$, $B=\multiset{0, 2, 2, 6}$, and  $$M=A\mink B = 
\multiset{0, 1, 2, 2, 3, 3, 3, 3, 5, 5, 5, 5, 6, 7, 9, 9}$$
%\medbreak
%\noindent
Assume  to run \proc{NewInitialSolution($M,C$)}. Where $C=\multiset{0, 2, 2, 5}$ is  a candidate 
solution but not a solution.
The matrix $mat$, if computed, would have  the form
$$
mat= \left[
\begin{matrix}
0&2&2&5\\
1&3&3&6\\
3&5&5&0\\
0&0&0&0
\end{matrix}
\right]
$$
\proc{NewInitialSolution($M,C$)} stops at line 20 returning $\multiset{0,2,2,5,1,3}$, i.e., the union of the first
row of $mat$ and the initial part (first 3 elements) of the first column of $mat$.
Experimental results clearly show that solutions to the problem contains with high probability elements
placed in the first row or in the first column of the matrix $mat$ associated to 
the local optimum candidate solution.

%=======================================================================
%=======================================================================
%=======================================================================
%=======================================================================
\section{Experimental results}\label{sec2.7}
%=======================================================================
%=======================================================================
%=======================================================================
%=======================================================================

We tested our algorithm 
on an 
iMac equipped 
with a $4.2$ GHz Intel Core $i7$ quad-core processor and  $32$ GB RAM ($2400$ MHz DDR4 ).
Operating System: macOS Monterey Version 12.2.1.
Our algorithm has been implemented in Wolfram Mathematica language (Version 12).
To make the code more readable even to those unfamiliar with the Mathematica language, 
we decided to describe it providing a pseudocode version (see Section \ref{sec3}).

%\medbreak%\noindent
Our algorithm has been extensively tested over instances (multisets of non-negative integers) 
of different size and structure. 
Instances depend on two parameters, namely $structure$ and $range$, 
and have been  generated according to the 
following procedure.

\newcommand{\Gen}{
\begin{codebox}
\Procname{$\proc{InstanceGeneration}(structure,range)$}     
\li $inst \gets \multiset{0}$
  \li 	\For $i \gets 1 $ \To $\attrib{structure}{length}$  		\Do  
  \li 	  Let $M$ be a multiset with the following properties:	 
  \li - cardinality of $M$ is equal to $structure[i]$
  \li - $M$ contains at least one element equal to $0$
   \li - each element of $M$ is randomly chosen in the interval $[0 \twodots range]$
 \li $inst \gets inst \mink M$
 \End
 \li \Return $inst$
 \end{codebox}
}

\Gen

The parameter $structure$ is a list of positive integers representing the cardinalities of the multisets
that, once summed together,  produce the instance.
The parameter $range$ represents an upper bound on the 
numbers in the multisets 
(see line $6$ of \proc{InstanceGeneration}).
As an example, the instance produced by  $\proc{InstanceGeneration}($\{2,2,3\}$,10)$
is a multiset of cardinality $12=2\times 2\times 3$ obtained by summing up 3 randomly generated 
multisets of cardinality $2,2$ and 3, respectively. Each element of the 3 multisets is randomly chosen
from the set $\{0,1,\dots,10\}$.
We only consider multisets containing at least one element equal to zero. In fact,
any multiset $M$ that does not contain 0, i.e., $\mu(0,M)=0$, can be always 
decomposed as $\multiset{min(M)} \mink M'$ where $M'$ is a multiset obtained from $M$ 
subtracting   to each element $min(M)$.
As an example, $\multiset{2,4,3,4,3,5}=\multiset{2}\mink \multiset{0,2,1,2,1,3}$.

%\medbreak%\noindent
For each $structure$ and $range$, we  tested our algorithms on a large number of instances
collecting results in Tables \ref{t1} to \ref{t12} in  Appendix \ref{data}. 

%\medbreak%\noindent
Columns of Tables contain the following data. \\
1. $Size$: size of the input, i.e., cardinality of the considered multiset \\
2. $Structure$: structure of the considered multiset \\
3. $Success$: percentage of runs for which a  solution is found\\
4. $Iterations$: Average number of iterations for any given  $structure$\\
5. $Time$: Average running time for any given  $structure$ \\
6. $Time/Iter$: $Time$ divided by $Iterations$ \\
7. $Time/Size$: $Time$ divided by $Size$
%\medbreak%\noindent
We investigated the performance of our algorithm in 
different scenarios.
 %\medbreak%\noindent
 Number of duplicates. We tested our algorithm
with two different values of the parameter $range$. Namely,
$range=5$ and $range=10000$. 
In the case of $range=5$, multisets contain a large number of duplicates, while in the case 
of $range=10000$ duplicates are very rare.

%\medbreak%\noindent  Type of structure. We tested our algorithm
with 3 different type of  structures $\{n,n\}$, $\{2\,n, n\}$ and $\{n,\dots,n\}$. \\
- $\{n,n\}$:  sum of two multisets with the same  cardinality;\\
- $\{2\,n,n\}$: sum of two multisets  with different cardinalities (one half of the other);  \\
- $\{n,\dots,n\}$:   sum of $k$ multisets with the same cardinality (denoted by $\{n\}^k$).

%\medbreak%\noindent
We now give  some  reading keys and interpretations  
of experimental data collected in Tables \ref{t1} to 
\ref{t12} in Appendix \ref{data}.

%\medbreak%\noindent
 \proc{IteratedSearch} finds a solution most of the time. Leaving unbounded the maximum number of 
allowed iterations, \proc{IteratedSearch}  always finds a solution. From a practical point of view, leaving 
unbounded the number of iterations prevents the algorithm to recognize irreducible multisets. 
In our tests we set the maximum number of iterations equal to 100. Even in this case, \proc{IteratedSearch}  is able to find a solution approximately $999$ times out of 1000.

%\medbreak%\noindent
Multisets with many duplicates approximately 
takes  the same amount of  time to decompose
with respect to multisets with a small number of duplicates. 
The presence of many duplicates forces the heuristics to go through a
 larger number of iterations  to find a solution but
 single iterations are much faster. 
With many duplicates, the behavior of \proc{IteratedSearch} is less regular in terms of 
running times and distribution of failures. 

%\medbreak%\noindent
Multisets obtained summing up many small multisets are much easier to decompose with respect to 
multisets obtained summing up 2 large multisets.
As an example, a multiset with structure \structureb{2}{15} and size 32768
 takes approximately the same time (last row of Table \ref{t3}) of a multiset with structure \structure{20,20} and size 400 (last row of Table \ref{t2}).
 For multisets obtained summing up many small multisets, the average number of iterations 
 is very close to 1.
  
%=======================================================================
%=======================================================================
%=======================================================================
%=======================================================================
\section{Polynomial Factorization vs  Iterated Search }\label{sec4.5}
%=======================================================================
%=======================================================================
%=======================================================================
%=======================================================================

 An alternative strategy for decomposing a multiset of non-negative integers
(or, equivalently, an intuitive way of factoring a polynomial in $\N[x]$)  might be the following.

\newcommand{\Astrategy}{
\begin{codebox}
\Procname{\proc{AlternativeStrategy}($M$)}   
\li \Comment $M$ is a multiset  of non-negative integers
\li $\id{p} \gets \proc{Polynomial}(M)$
\li $\id{fl} \gets \proc{FactorList}(p)$
\li $\id{(P_1,P_2)} \gets \proc{Group}(\id{fl})$
\li   \Return $(Multiset(P_1),Multiset(P_2))$
\end{codebox}
}

\Astrategy

%\noindent
Line 2 computes the polynomial $p$ associated to the multiset $M$ as shown in Equation (\ref{pol}).
Line 3 computes the factor list $fl$ of $p$. 
Line 4, using some unknown algorithm  (it would be of some interest to find an algorithm for efficiently computing $\proc{Group}(\id{fl}))$, computes a partition $P=\{P_1,P_2\}$ (if there exists one)  of the factor list $\id{fl}$ such that the product
of all the polynomials in  $P_1$ and the product
of all the polynomials in  $P_2$ have non-negative coefficients.

In what follows we will assume that the computational cost of Line 4 is zero.
Table \ref{t13} to \ref{t15} compare  running times of  \proc{IteratedSearch} and \proc{AlternativeStrategy}
for multisets with homogeneous $structure$ and increasing $ranges$.

For computing the factor list at Line 3 of \proc{AlternativeStrategy} we 
make use of the function \proc{FactorList} provided by Mathematica Language
(similar results are obtained by using the function \proc{factor} of MatLab).

Experimental results (see Tables \ref{t13},\ref{t14} and \ref{t15}) clearly show that the running time of \proc{IteratedSearch} is independent of the magnitude of numbers in the multisets (exponents in the polynomials).
\proc{IteratedSearch}  is much faster than \proc{AlternativeStrategy}
in the case of multisets containing large numbers and small multiplicity.

Doing the reverse path enable us to give a new technique for decomposing polynomials in $\N[x]$
based on \proc{IteratedSearch}.

\newcommand{\Bstrategy}{
\begin{codebox}
\Procname{\proc{$\N$-PolyFact}($p$)}   
\li \Comment $p\in \N[x]$ 
\li $\id{M} \gets \proc{Multiset}(p)$
\li $\id{S} \gets\proc{IteratedSearch(M)}$ \Comment 
\li $\id{P} \gets Polynomial(S)$
\li   \Return $(Polynomial(S),p/P)$
\end{codebox}
}

\Bstrategy

We end this section by giving 
%an example of 
a  small multiset $M$ of non-negative integers
that $\proc{IteratedSearch}$ decomposes in $0.008$ seconds. 
$\proc{AlternativeStrategy}$ (both using Mathematica and MatLab factorization primitives) called on the same multiset, after 24 
hours of computation, was unable to find any solution.

$$A=\{\!\{0, 1249, 4270, 4324, 4852\}\!\}$$
$$B=\{\!\{0, 1705, 2250, 2267, 4390\}\!\}$$

\begin{eqnarray*}
M=A\mink B&=&\{\!\{0, 1249, 1705, 2250, 2267, 2954, 3499, 3516, 4270, 4324, 4390, 4852,5639,\\
&&  5975, 6029, 6520, 6537, 6557, 6574, 6591, 7102, 7119, 8660, 
8714, 9242\}\!\}
\end{eqnarray*}
 \begin{eqnarray*}
Polynomial(M)&=&1 + x^{1249} + x^{1705} + x^{2250} + x^{2267} + x^{2954} + x^{3499} +x^{3516} + x^{4270} +\\
&&   x^{4324} + x^{4390} +x^{4852} + x^{5639} + x^{5975} + x^{6029} + x^{6520} +x^{6537} +  \\
&&  x^{6557} + x^{6574} + x^{6591} + x^{7102} + x^{7119} + x^{8660}+ x^{8714} + x^{9242}
\end{eqnarray*}

%=======================================================================
%=======================================================================
%=======================================================================
%=======================================================================
\section{Conclusions and further work}\label{sec5}
%=======================================================================
%=======================================================================
%=======================================================================
%=======================================================================

We have introduced and analyzed a heuristic technique for decomposing multisets of non-negative integers according to the Minkowski sum.
Experimental results show that our technique allows to decompose quite 
large multisets (hundreds to thousands of elements depending on the instance structure)  
in seconds.
Our technique can also be used to tackle the 
problem of  factoring polynomials in $\N[x]$. 
Experimental results show that, when the size of exponents (elements of  multisets)
 increases, our technique is much faster than state-of-the-art implementation of 
polynomial factoring algorithms over $\Z[x]$ that can be viewed as a preparatory step for factoring 
over $\N[x]$. 

A natural extension of this work is replacing non-negative integers 
with more complex mathematical objects. 
It would be of some interest to investigate the case of $d$ dimensional vectors of non-negative 
integers with $d>1$. The problem of decomposing multisets of $d$ dimensional vectors is strictly related 
to the problem of factoring multivariate polynomials with non-negative coefficients, but also
to a number of problems arising, for example, in the field of computational geometry and seems to be  more challenging than the 1 dimensional case. 

It would be interesting to investigate whether the combination of the results obtained by using 
our algorithm on single components of the $d$ dimensional object can be of any help for solving the global problem.

\clearpage

\appendix

\section{Experimental Data Tables I}\label{data}

\begin{table}[h!]
\caption{$Range=5$. Number of tested instances for each structure: $1000$.
 }
\begin{center}
\begin{tabular}{|c|c|c|c|c|c|c|}
\hline
 Size&Structure& Success & Iterations& Time & Time/Iter & Time/Size\\
\hline
9&\structure{3,3}&100&1&0.001&0.001&0.00011\\\hline
16&\structure{4,4}&100&1.14&0.002&0.00175&0.00012\\\hline
25&\structure{5,5}&100&1.4&0.008&0.00571&0.00032\\\hline
36&\structure{6,6}&100&1.9&0.025&0.01316&0.00069\\\hline
49&\structure{7,7}&100&2.6&0.061&0.02346&0.00124\\\hline
64&\structure{8,8}&100&3&0.122&0.04067&0.00191\\\hline
81&\structure{9,9}&100&4.22&0.242&0.05735&0.00299\\\hline
100&\structure{10,10}&100&4.22&0.366&0.08673&0.00366\\\hline
121&\structure{11,11}&100&7.06&0.934&0.13229&0.00772\\\hline
144&\structure{12,12}&100&4.72&0.95&0.20127&0.0066\\\hline
169&\structure{13,13}&100&11.7&2.728&0.23316&0.01614\\\hline
196&\structure{14,14}&100&7.02&2.454&0.34957&0.01252\\\hline
225&\structure{15,15}&100&7.14&3.298&0.4619&0.01466\\\hline
256&\structure{16,16}&100&8.16&4.563&0.55919&0.01782\\\hline
289&\structure{17,17}&100&10.72&8.151&0.76035&0.0282\\\hline
324&\structure{18,18}&100&9.56&9.18&0.96025&0.02833\\\hline
361&\structure{19,19}&99.9&11&12.168&1.10618&0.03371\\\hline
400&\structure{20,20}&100&18.5&29.491&1.59411&0.07373\\\hline

\end{tabular}
\end{center}

\label{t1}
\end{table}

%-------------------

\begin{table}[h!]
\caption{$Range=5$. Number of tested instances for each structure: $1000$.}
 
\begin{center}
\begin{tabular}{|c|c|c|c|c|c|c|}
\hline
 Size&Structure& Success & Iterations& Time & Time/Iter & Time/Size\\
\hline
18&\structure{6,3}&100&1.24&0.002&0.00161&0.00011\\\hline
32&\structure{8,4}&100&1.92&0.011&0.00573&0.00034\\\hline
50&\structure{10,5}&100&2.52&0.038&0.01508&0.00076\\\hline
72&\structure{12,6}&100&3.1&0.104&0.03355&0.00144\\\hline
98&\structure{14,7}&100&3.58&0.222&0.06201&0.00227\\\hline
128&\structure{16,8}&100&4.68&0.452&0.09658&0.00353\\\hline
162&\structure{18,9}&100&6.44&0.961&0.14922&0.00593\\\hline
200&\structure{20,10}&100&9.22&1.865&0.20228&0.00932\\\hline
242&\structure{22,11}&100&7.56&2.655&0.35119&0.01097\\\hline
288&\structure{24,12}&100&9.4&4.161&0.44266&0.01445\\\hline
338&\structure{26,13}&100&15.8&8.474&0.53633&0.02507\\\hline
392&\structure{28,14}&99.9&12.1&10.56&0.87273&0.02694\\\hline
450&\structure{30,15}&100&11.62&13.641&1.17392&0.03031\\\hline

\end{tabular}
\end{center}

\label{t2}
\end{table}

%-------------------

\begin{table}[h!]
\caption{$Range=5$. Number of tested instances for each structure: $1000$.}
\begin{center}
\begin{tabular}{|c|c|c|c|c|c|c|}
\hline
 Size&Structure& Success & Iterations& Time & Time/Iter & Time/Size\\
\hline
8&\structureb{2}{3}&100&1&0.001&0.001&0.00012\\\hline
16&\structureb{2}{4}&100&1&0.001&0.001&0.00006\\\hline
32&\structureb{2}{5}&100&1&0.001&0.001&0.00003\\\hline
64&\structureb{2}{6}&100&1&0.002&0.002&0.00003\\\hline
128&\structureb{2}{7}&100&1&0.004&0.004&0.00003\\\hline
256&\structureb{2}{8}&100&1&0.007&0.007&0.00003\\\hline
512&\structureb{2}{9}&100&1&0.014&0.014&0.00003\\\hline
1024&\structureb{2}{10}&100&1&0.037&0.037&0.00004\\\hline
2048&\structureb{2}{11}&100&1&0.11&0.11&0.00005\\\hline
4096&\structureb{2}{12}&100&1&0.375&0.375&0.00009\\\hline
8192&\structureb{2}{13}&100&1&1.15&1.15&0.00014\\\hline
16384&\structureb{2}{14}&100&1&4.708&4.708&0.00029\\\hline
32768&\structureb{2}{15}&100&1&18.625&18.625&0.00057\\\hline

\end{tabular}
\end{center}

\label{t3}
\end{table}

%-------------------

\begin{table}[h!]
\caption{$Range=5$. Number of tested instances for each structure: $1000$.}
\begin{center}
\begin{tabular}{|c|c|c|c|c|c|c|}
\hline
 Size&Structure& Success & Iterations& Time & Time/Iter & Time/Size\\
\hline
27&\structureb{3}{3}&100&1&0.003&0.003&0.00011\\\hline
81&\structureb{3}{4}&100&1.04&0.012&0.01154&0.00015\\\hline
243&\structureb{3}{5}&100&1&0.039&0.039&0.00016\\\hline
729&\structureb{3}{6}&100&1&0.175&0.175&0.00024\\\hline
2187&\structureb{3}{7}&100&1&1.088&1.088&0.0005\\\hline
6561&\structureb{3}{8}&100&1&6.646&6.646&0.00101\\\hline
19683&\structureb{3}{9}&100&1&60.155&60.155&0.00306\\\hline

\end{tabular}
\end{center}

\label{t4}
\end{table}

%-------------------

\begin{table}[h!]
\caption{$Range=5$. Number of tested instances for each structure: $1000$.
For $Size=16384$, due to time limits,  we reduced the number of instances to $300$.
}
\begin{center}
\begin{tabular}{|c|c|c|c|c|c|c|}
\hline
 Size&Structure& Success & Iterations& Time & Time/Iter & Time/Size\\
\hline
64&\structureb{4}{3}&100&1.1&0.022&0.02&0.00034\\\hline
256&\structureb{4}{4}&100&1.02&0.14&0.13725&0.00055\\\hline
1024&\structureb{4}{5}&100&1.06&1.266&1.19434&0.00124\\\hline
4096&\structureb{4}{6}&100&1&11.377&11.377&0.00278\\\hline
16384&\structureb{4}{7}&100&1.24&366.325&295.423&0.02236\\\hline

\end{tabular}
\end{center}

\label{t5}
\end{table}

%-------------------

\begin{table}[h!]
\caption{$Range=5$. Number of tested instances for each structure: $1000$.
For $Size=15625$, due to time limits,  we reduced the number of instances to $300$.
}
\begin{center}
\begin{tabular}{|c|c|c|c|c|c|c|}
\hline
 Size&Structure& Success & Iterations& Time & Time/Iter & Time/Size\\
\hline
125&\structureb{5}{3}&100&1.38&0.114&0.08261&0.00091\\\hline
625&\structureb{5}{4}&100&1.26&1.307&1.0373&0.00209\\\hline
3125&\structureb{5}{5}&100&1.12&23.818&21.2661&0.00762\\\hline
15625&\structureb{5}{6}&100&1.08&521.383&482.762&0.03337\\\hline

\end{tabular}
\end{center}

\label{t6}
\end{table}

%-------------------

\begin{table}[h!]
\caption{$Range=10000$.  Number of instances for each structure: $1000$.
 }
\begin{center}
\begin{tabular}{|c|c|c|c|c|c|c|}
\hline
 Size&Structure& Success & Iterations& Time & Time/Iter & Time/Size\\
\hline
9&\structure{3,3}&100&1&0.001&0.001&0.00011\\\hline
16&\structure{4,4}&100&1&0.002&0.002&0.00012\\\hline
25&\structure{5,5}&100&1&0.008&0.008&0.00032\\\hline
36&\structure{6,6}&100&1&0.02&0.02&0.00056\\\hline
49&\structure{7,7}&100&1&0.05&0.05&0.00102\\\hline
64&\structure{8,8}&100&1&0.105&0.105&0.00164\\\hline
81&\structure{9,9}&100&1&0.199&0.199&0.00246\\\hline
100&\structure{10,10}&100&1&0.375&0.375&0.00375\\\hline
121&\structure{11,11}&100&1&0.606&0.606&0.00501\\\hline
144&\structure{12,12}&100&1&1.138&1.138&0.0079\\\hline
169&\structure{13,13}&100&1&1.815&1.815&0.01074\\\hline
196&\structure{14,14}&100&1&2.831&2.831&0.01444\\\hline
225&\structure{15,15}&100&1&4.064&4.064&0.01806\\\hline
256&\structure{16,16}&100&1&6.09&6.09&0.02379\\\hline
289&\structure{17,17}&100&1.4&10.515&7.51071&0.03638\\\hline
324&\structure{18,18}&100&1&13.469&13.469&0.04157\\\hline
361&\structure{19,19}&100&1&19.217&19.217&0.05323\\\hline
400&\structure{20,20}&100&1.02&27.122&26.5902&0.0678\\\hline

\end{tabular}
\end{center}

\label{t7}
\end{table}

%-------------------

\begin{table}[h!]
\caption{$Range=10000$.  Number of instances for each structure: $1000$.
 }
\begin{center}
\begin{tabular}{|c|c|c|c|c|c|c|}
\hline
 Size&Structure& Success & Iterations& Time & Time/Iter & Time/Size\\
\hline
18&\structure{6,3}&100&2.22&0.004&0.0018&0.00022\\\hline
32&\structure{8,4}&100&1.78&0.012&0.00674&0.00038\\\hline
50&\structure{10,5}&100&2.2&0.047&0.02136&0.00094\\\hline
72&\structure{12,6}&100&1.76&0.096&0.05455&0.00133\\\hline
98&\structure{14,7}&100&1.72&0.214&0.12442&0.00218\\\hline
128&\structure{16,8}&100&1.96&0.488&0.24898&0.00381\\\hline
162&\structure{18,9}&100&3.02&1.469&0.48642&0.00907\\\hline
200&\structure{20,10}&99.9&6.14&6.141&100016&0.0307\\\hline
242&\structure{22,11}&100&2.68&3.944&1.47164&0.0163\\\hline
288&\structure{24,12}&100&1.64&4.777&2.9128&0.01659\\\hline
338&\structure{26,13}&100&2.26&9.864&4.3646&0.02918\\\hline
392&\structure{28,14}&100&2.06&15.012&7.28738&0.0383\\\hline
450&\structure{30,15}&100&3.18&33.11&10.412&0.07358\\\hline

\end{tabular}
\end{center}

\label{t8}
\end{table}

%-------------------

\begin{table}[h!]
\caption{$Range=10000$.  Number of instances for each structure: $1000$.
}
\begin{center}
\begin{tabular}{|c|c|c|c|c|c|c|}
\hline
 Size&Structure& Success & Iterations& Time & Time/Iter & Time/Size\\
\hline
8&\structureb{2}{3}&100&1&0.001&0.001&0.00012\\\hline
16&\structureb{2}{4}&100&1&0.001&0.001&0.00006\\\hline
32&\structureb{2}{5}&100&1&0.001&0.001&0.00003\\\hline
64&\structureb{2}{6}&100&1&0.001&0.001&0.00002\\\hline
128&\structureb{2}{7}&100&1&0.002&0.002&0.00002\\\hline
256&\structureb{2}{8}&100&1&0.004&0.004&0.00002\\\hline
512&\structureb{2}{9}&100&1&0.009&0.009&0.00002\\\hline
1024&\structureb{2}{10}&100&1&0.024&0.024&0.00002\\\hline
2048&\structureb{2}{11}&100&1&0.07&0.07&0.00003\\\hline
4096&\structureb{2}{12}&100&1&0.228&0.228&0.00006\\\hline
8192&\structureb{2}{13}&100&1&0.811&0.811&0.0001\\\hline
16384&\structureb{2}{14}&100&1&2.984&2.984&0.00018\\\hline
32768&\structureb{2}{15}&100&1&11.708&11.708&0.00036\\\hline

\end{tabular}
\end{center}

\label{t9}
\end{table}

%-------------------

\begin{table}[h!]
\caption{$Range=10000$.  Number of instances for each structure: $1000$.
 }
\begin{center}
\begin{tabular}{|c|c|c|c|c|c|c|}
\hline
 Size&Structure& Success & Iterations& Time & Time/Iter & Time/Size\\
\hline
27&\structureb{3}{3}&100&1&0.002&0.002&0.00007\\\hline
81&\structureb{3}{4}&100&1&0.008&0.008&0.0001\\\hline
243&\structureb{3}{5}&100&1&0.028&0.028&0.00012\\\hline
729&\structureb{3}{6}&100&1&0.139&0.139&0.00019\\\hline
2187&\structureb{3}{7}&100&1&0.916&0.916&0.00042\\\hline
6561&\structureb{3}{8}&100&1&7.047&7.047&0.00107\\\hline
19683&\structureb{3}{9}&100&1&72.214&72.214&0.00367\\\hline

\end{tabular}
\end{center}

\label{t10}
\end{table}

%-------------------

\begin{table}[h!]
\caption{$Range=10000$.  Number of instances for each structure: $1000$.
For $Size=16384$, due to time limits,  we reduced the number of instances to $300$.
 }
\begin{center}
\begin{tabular}{|c|c|c|c|c|c|c|}
\hline
 Size&Structure& Success & Iterations& Time & Time/Iter & Time/Size\\
\hline
64&\structureb{4}{3}&100&1.16&0.021&0.0181&0.00033\\\hline
256&\structureb{4}{4}&100&1.12&0.148&0.13214&0.00058\\\hline
1024&\structureb{4}{5}&100&1.28&1.54&1.20312&0.0015\\\hline
4096&\structureb{4}{6}&100&1.26&21.532&17.0889&0.00526\\\hline
16384&\structureb{4}{7}&100&1.18&355.661&301.408&0.02171\\\hline

\end{tabular}
\end{center}

\label{t11}
\end{table}

%-------------------

\begin{table}[h!]
\caption{$Range=10000$.  Number of instances for each structure: $1000$.
 For $Size=3125$ and $Size=15625$, due to time limits,  we reduced the number of instances to $100$.
}
\begin{center}
\begin{tabular}{|c|c|c|c|c|c|c|}
\hline
 Size&Structure& Success & Iterations& Time & Time/Iter & Time/Size\\
\hline
125&\structureb{5}{3}&100&1.62&0.146&0.09012&0.00117\\\hline
625&\structureb{5}{4}&100&5.2&4.767&0.91673&0.00763\\\hline
3125&\structureb{5}{5}&100&7.7&122.143&15.8627&0.03909\\\hline
15625&\structureb{5}{6}&100&3.4&1689.16&496.812&0.10811\\\hline

\end{tabular}
\end{center}

\label{t12}
\end{table}

%-------------------

\clearpage

\section{Experimental Data Tables II}\label{data2}

\begin{table}[h!]
\caption{Running times for \proc{IteratedSearch} and \proc{AlternativeStrategy}
called on multisets with different $range$ values and $structure=$ \structure{5,5}.  
Number of instances for each range: $100$.
 }
\begin{center}
\begin{tabular}{|c|c|c|c|c|}
\hline
 Size&Structure& Range & \proc{IteratedSearch} & \proc{AlternativeStrategy} \\
\hline
25&\structure{5,5}&$100$&0.09&0.144747\\\hline
25&\structure{5,5}&$300$&0.008&3.764507\\\hline
25&\structure{5,5}&$500$&0.009&22.003455\\\hline
25&\structure{5,5}&$700$&0.008&64.317906\\\hline
25&\structure{5,5}&$900$&0.007&161.541679\\\hline
25&\structure{5,5}&$1100$&0.01&253.745332\\\hline
\end{tabular}
\end{center}

\label{t13}
\end{table}

\begin{table}[h!]
\caption{Running times for \proc{IteratedSearch} and \proc{AlternativeStrategy}
called on multisets with different $range$ values and $structure=$ \structure{10,10}.  
Number of instances for each range: $100$.
 }
\begin{center}
\begin{tabular}{|c|c|c|c|c|}
\hline
 Size&Structure& Range & \proc{IteratedSearch} & \proc{AlternativeStrategy} \\
\hline
100&\structure{10,10}&$100$&0.551&0.220502\\\hline
100&\structure{10,10}&$300$&0.532&4.213079\\\hline
100&\structure{10,10}&$500$&0.397&26.706801\\\hline
100&\structure{10,10}&$700$&0.426&75.783461\\\hline
100&\structure{10,10}&$900$&0.612&187.938575\\\hline
100&\structure{10,10}&$1100$&0.4&379.374113\\\hline
\end{tabular}
\end{center}

\label{t14}
\end{table}

\begin{table}[h!]
\caption{Running times for \proc{IteratedSearch} and \proc{AlternativeStrategy}
called on multisets with different $range$ values and $structure=$ \structureb{2}{12}.  
Number of instances for each range: $100$.
 }
\begin{center}
\begin{tabular}{|c|c|c|c|c|}
\hline
 Size&Structure& Range & \proc{IteratedSearch} & \proc{AlternativeStrategy} \\
\hline
4096&\structureb{2}{12}&$40$&0.294&0.142596\\\hline
4096&\structureb{2}{12}&$60$&0.319&2.682864\\\hline
4096&\structureb{2}{12}&$80$&0.316&3.145838\\\hline
4096&\structureb{2}{12}&$100$&0.311&6.137466\\\hline
4096&\structureb{2}{12}&$120$&0.283&31.849028\\\hline
4096&\structureb{2}{12}&$140$&0.253&356.950613\\\hline
\end{tabular}
\end{center}

\label{t15}
\end{table}

%=======================================================================
%=======================================================================
%=======================================================================
%=======================================================================
%=======================================================================
%=======================================================================
%=======================================================================

\clearpage
\nocite{*}
\bibliographystyle{abbrvnat}
% use the following instead if you encounter problems 
%\bibliographystyle{alpha}
\bibliography{sample-dmtcs}

\begin{thebibliography}{36}
\providecommand{\natexlab}[1]{#1}
\providecommand{\url}[1]{\texttt{#1}}
\expandafter\ifx\csname urlstyle\endcsname\relax
  \providecommand{\doi}[1]{doi: #1}\else
  \providecommand{\doi}{doi: \begingroup \urlstyle{rm}\Url}\fi

\bibitem[Anderson(1989)]{AndersonIan1989Cofs}
I.~Anderson.
\newblock \emph{Combinatorics of finite sets}.
\newblock Oxford science publications. lat. Clarendon Press ; Oxford University
  Press, Oxford [England] : New York, 1989.
\newblock ISBN 0198533799.

\bibitem[Anderson(2002)]{anderson2002combinatorics}
I.~Anderson.
\newblock \emph{Combinatorics of Finite Sets}.
\newblock Dover books on mathematics. Dover Publications, 2002.
\newblock ISBN 9780486422572.
\newblock URL \url{https://books.google.it/books?id=RjDd4RaqrIwC}.

\bibitem[Berthaud et~al.(2014)Berthaud, Capelli, Gustedt, Kirchner, Loiseau,
  Magron, Medves, Monteil, Riverieux, and Romary]{berthaud:hal-01002815}
C.~Berthaud, L.~Capelli, J.~Gustedt, C.~Kirchner, K.~Loiseau, A.~Magron,
  M.~Medves, A.~Monteil, G.~Riverieux, and L.~Romary.
\newblock {EPISCIENCES - an overlay publication platform}.
\newblock In D.~P. Polydoratou, editor, \emph{{ELPUB2014 - International
  Conference on Electronic Publishing}}, pages 78--87, Thessalonique, Greece,
  June 2014. {Alexander Technological Education Institute of Thessaloniki},
  {IOS Press}.
\newblock \doi{10.3233/978-1-61499-409-1-78}.
\newblock URL \url{https://hal.inria.fr/hal-01002815}.

\bibitem[Brunotte(2013)]{Brunotte}
H.~Brunotte.
\newblock On some classes of polynomials with nonnegative coefficients and a
  given factor.
\newblock \emph{Periodica Mathematica Hungarica}, 67\penalty0 (1):\penalty0
  15--32, 2013.
\newblock \doi{10.1007/s10998-013-2367-8}.
\newblock URL \url{https://doi.org/10.1007/s10998-013-2367-8}.

\bibitem[Campanini and Facchini(2019)]{fac19}
F.~Campanini and A.~Facchini.
\newblock Factorizations of polynomials with integral non-negative
  coefficients.
\newblock \emph{Semigroup Forum}, 99\penalty0 (2):\penalty0 317--332, 2019.
\newblock \doi{10.1007/s00233-018-9979-5}.
\newblock URL \url{https://doi.org/10.1007/s00233-018-9979-5}.

\bibitem[Cormen et~al.(2009)Cormen, Leiserson, Rivest, and Stein]{cormen}
T.~H. Cormen, C.~E. Leiserson, R.~L. Rivest, and C.~Stein.
\newblock \emph{Introduction to Algorithms}.
\newblock The MIT Press, 3rd edition, 2009.

\bibitem[Cucker et~al.(1999)Cucker, Koiran, and Smale]{CUCKER199921}
F.~Cucker, P.~Koiran, and S.~Smale.
\newblock A polynomial time algorithm for diophantine equations in one
  variable.
\newblock \emph{Journal of Symbolic Computation}, 27\penalty0 (1):\penalty0
  21--29, 1999.
\newblock ISSN 0747-7171.
\newblock \doi{https://doi.org/10.1006/jsco.1998.0242}.
\newblock URL
  \url{https://www.sciencedirect.com/science/article/pii/S0747717198902425}.

\bibitem[DeVos et~al.(2013)DeVos, Krakovski, Mohar, and
  Sheikh~Ahmady]{DeVos2013}
M.~DeVos, R.~Krakovski, B.~Mohar, and A.~Sheikh~Ahmady.
\newblock Integral cayley multigraphs over abelian and hamiltonian groups.
\newblock \emph{The Electronic Journal of Combinatorics}, 20\penalty0 (2), Jun
  2013.
\newblock ISSN 1077-8926.
\newblock \doi{10.37236/2742}.
\newblock URL \url{http://dx.doi.org/10.37236/2742}.

\bibitem[Dorwart(1935)]{Dorwart}
H.~L. Dorwart.
\newblock Irreducibility of polynomials.
\newblock \emph{The American Mathematical Monthly}, 42\penalty0 (6):\penalty0
  369--381, 1935.
\newblock \doi{10.1080/00029890.1935.11987732}.

\bibitem[Dudek et~al.(2013)Dudek, Frieze, Ruci\'{n}ski, and
  \v{S}ileikis]{DUDEK2013785}
A.~Dudek, A.~Frieze, A.~Ruci\'{n}ski, and M.~\v{S}ileikis.
\newblock Approximate counting of regular hypergraphs.
\newblock \emph{Information Processing Letters}, 113\penalty0 (19):\penalty0
  785--788, 2013.
\newblock ISSN 0020-0190.
\newblock \doi{https://doi.org/10.1016/j.ipl.2013.07.018}.
\newblock URL
  \url{https://www.sciencedirect.com/science/article/pii/S002001901300207X}.

\bibitem[Emiris et~al.(2017)Emiris, Karasoulou, and Tzovas]{Ioa17}
I.~Z. Emiris, A.~Karasoulou, and C.~Tzovas.
\newblock Approximating multidimensional subset sum and minkowski decomposition
  of polygons.
\newblock \emph{Mathematics in Computer Science}, 11\penalty0 (1):\penalty0
  35--48, 2017.
\newblock \doi{10.1007/s11786-017-0297-1}.
\newblock URL \url{https://doi.org/10.1007/s11786-017-0297-1}.

\bibitem[Fellows and Koblitz(1993)]{FellowsK93}
M.~R. Fellows and N.~Koblitz.
\newblock Fixed-parameter complexity and cryptography.
\newblock In G.~D. Cohen, T.~Mora, and O.~Moreno, editors, \emph{Applied
  algebra, algebraic algorithms and error-correcting codes, 10th International
  Symposium, AAECC-10, San Juan de Puerto Rico, Puerto Rico, May 10-14, 1993,
  Proceedings}, volume 673 of \emph{Lecture Notes in Computer Science}, pages
  121--131. Springer, 1993.
\newblock ISBN 3-540-56686-4.

\bibitem[Gao and Lauder(2001)]{Gao01}
S.~Gao and A.~G.~B. Lauder.
\newblock Decomposition of polytopes and polynomials.
\newblock \emph{Discrete \& Computational Geometry}, 26\penalty0 (1):\penalty0
  89--104, 2001.
\newblock \doi{10.1007/s00454-001-0024-0}.
\newblock URL \url{https://doi.org/10.1007/s00454-001-0024-0}.

\bibitem[Garey and Johnson(1979)]{garey1979computers}
M.~R. Garey and D.~S. Johnson.
\newblock \emph{Computers and intractability: a guide to the theory of
  NP-Completeness (Series of Books in the Mathematical Sciences)}.
\newblock W. H. Freeman, 1979.
\newblock ISBN 0716710455.
\newblock URL
  \url{http://www.amazon.com/Computers-Intractability-NP-Completeness-Mathematical-Sciences/dp/0716710455}.

\bibitem[Garey and S.Johnson(1978)]{GJ78}
M.~R. Garey and D.~S.Johnson.
\newblock ``strong'' np-completeness results: motivation, examples, and
  iimplications.
\newblock \emph{J. ACM}, 25\penalty0 (3):\penalty0 499--508, jul 1978.
\newblock ISSN 0004-5411.
\newblock \doi{10.1145/322077.322090}.
\newblock URL \url{https://doi.org/10.1145/322077.322090}.

\bibitem[Grenet(2016)]{GRENET2016171}
B.~Grenet.
\newblock Bounded-degree factors of lacunary multivariate polynomials.
\newblock \emph{Journal of Symbolic Computation}, 75:\penalty0 171--192, 2016.
\newblock ISSN 0747-7171.
\newblock \doi{https://doi.org/10.1016/j.jsc.2015.11.013}.
\newblock Special issue on the conference ISSAC 2014: Symbolic computation and
  computer algebra.

\bibitem[Grumbach and Milo(1996)]{GrumbachM96}
S.~Grumbach and T.~Milo.
\newblock Towards tractable algebras for bags.
\newblock \emph{J. Comput. Syst. Sci.}, 52\penalty0 (3):\penalty0 570--588,
  1996.
\newblock \doi{10.1006/jcss.1996.0042}.
\newblock URL \url{https://doi.org/10.1006/jcss.1996.0042}.

\bibitem[Henglein et~al.(2022)Henglein, Kaarsgaard, and Mathiesen]{friz22}
F.~Henglein, R.~Kaarsgaard, and M.~K. Mathiesen.
\newblock The programming of algebra.
\newblock \emph{CoRR}, abs/2207.00850, 2022.
\newblock \doi{10.48550/arXiv.2207.00850}.
\newblock URL \url{https://doi.org/10.48550/arXiv.2207.00850}.

\bibitem[Hoeij(2002)]{VANHOEIJ2002167}
M.~V. Hoeij.
\newblock Factoring polynomials and the knapsack problem.
\newblock \emph{Journal of Number Theory}, 95\penalty0 (2):\penalty0 167--189,
  2002.
\newblock ISSN 0022-314X.
\newblock \doi{https://doi.org/10.1006/jnth.2001.2763}.
\newblock URL
  \url{https://www.sciencedirect.com/science/article/pii/S0022314X01927635}.

\bibitem[Johnson(1981)]{JOHNSON1981393}
D.~S. Johnson.
\newblock The np-completeness column: An ongoing guide.
\newblock \emph{Journal of Algorithms}, 2\penalty0 (4):\penalty0 393--405,
  1981.
\newblock ISSN 0196-6774.
\newblock \doi{https://doi.org/10.1016/0196-6774(81)90037-7}.
\newblock URL
  \url{https://www.sciencedirect.com/science/article/pii/0196677481900377}.

\bibitem[Kaltofen(1992)]{DBLP:conf/latin/Kaltofen92}
E.~Kaltofen.
\newblock Polynomial factorization 1987-1991.
\newblock In I.~Simon, editor, \emph{{LATIN} '92, 1st Latin American Symposium
  on Theoretical Informatics, S{\~{a}}o Paulo, Brazil, April 6-10, 1992,
  Proceedings}, volume 583 of \emph{Lecture Notes in Computer Science}, pages
  294--313. Springer, 1992.
\newblock \doi{10.1007/BFb0023837}.
\newblock URL \url{https://doi.org/10.1007/BFb0023837}.

\bibitem[Kaltofen and Koiran(2005)]{KaltofenK05}
E.~Kaltofen and P.~Koiran.
\newblock On the complexity of factoring bivariate supersparse (lacunary)
  polynomials.
\newblock In M.~Kauers, editor, \emph{ISSAC}, pages 208--215. ACM, 2005.
\newblock ISBN 1-59593-095-7.

\bibitem[Karp(1972)]{Kar72}
R.~Karp.
\newblock Reducibility among combinatorial problems.
\newblock In R.~Miller and J.~Thatcher, editors, \emph{Complexity of computer
  computations}, pages 85--103. Plenum Press, 1972.

\bibitem[Karpinski and Shparlinski(1999)]{KarpinskiS99}
M.~Karpinski and I.~E. Shparlinski.
\newblock On the computational hardness of testing square-freeness of sparse
  polynomials.
\newblock In M.~P.~C. Fossorier, H.~Imai, S.~Lin, and A.~Poli, editors,
  \emph{AAECC}, volume 1719 of \emph{Lecture Notes in Computer Science}, pages
  492--497. Springer, 1999.
\newblock ISBN 3-540-66723-7.

\bibitem[Kim and Roush(2005)]{Kim05}
K.~H. Kim and F.~W. Roush.
\newblock Factorization of polynomials in one variable over the tropical
  semiring.
\newblock \emph{https://arxiv.org/abs/math/0501167}, 2005.
\newblock \doi{https://doi.org/10.48550/arXiv.math/0501167}.
\newblock URL \url{https://arxiv.org/abs/math/0501167}.

\bibitem[Lamperti et~al.(2000)Lamperti, Melchiori, and Zanella]{lamp00}
G.~Lamperti, M.~Melchiori, and M.~Zanella.
\newblock On multisets in database systems.
\newblock volume 2235, pages 147--216, 08 2000.
\newblock ISBN 978-3-540-43063-6.
\newblock \doi{10.1007/3-540-45523-X_9}.

\bibitem[Lenstra et~al.(1982)Lenstra, Lenstra, and
  Lov\'asz]{Lenstra82factoringpolynomials}
A.~K. Lenstra, H.~W. Lenstra, and L.~Lov\'asz.
\newblock Factoring polynomials with rational coefficients.
\newblock \emph{Mathematische Annalen}, 261:\penalty0 515--534, 1982.

\bibitem[Ng et~al.(2010)Ng, Barketau, Cheng, and Kovalyov]{NG2010601}
C.~Ng, M.~Barketau, T.~Cheng, and M.~Y. Kovalyov.
\newblock Product partition and related problems of scheduling and systems
  reliability: computational complexity and approximation.
\newblock \emph{European Journal of Operational Research}, 207\penalty0
  (2):\penalty0 601--604, 2010.
\newblock ISSN 0377-2217.
\newblock \doi{https://doi.org/10.1016/j.ejor.2010.05.034}.
\newblock URL
  \url{https://www.sciencedirect.com/science/article/pii/S0377221710003905}.

\bibitem[Oetiker et~al.(1999)Oetiker, Partl, Hyna, and
  Schlegl]{oetiker99:_not_so_short_introd_latex}
T.~Oetiker, H.~Partl, I.~Hyna, and E.~Schlegl.
\newblock \emph{The Not So Short Introduction to \LaTeX2e}, 3.3 edition, 1999.
\newblock available at
  \href{http://www.loria.fr/services/tex/general/lshort2e.pdf}{\texttt{http://www.loria.fr/services/tex/general/lshort2e.pdf}}.

\bibitem[Plaisted(1977)]{Plaisted77}
D.~A. Plaisted.
\newblock Sparse complex polynomials and polynomial reducibility.
\newblock \emph{J. Comput. Syst. Sci.}, 14\penalty0 (2):\penalty0 210--221,
  1977.

\bibitem[Schubert(1793)]{Schubert}
F.~T. Schubert.
\newblock De inventione divisorum.
\newblock \emph{Nova Acta Academiae Scientiarum Petropolitanae}, 11:\penalty0
  172--182, 1793.

\bibitem[Singh et~al.(2007)Singh, Ibrahim, Yohanna, and Singh]{Singh2007}
D.~Singh, A.~M. Ibrahim, T.~Yohanna, and J.~N. Singh.
\newblock An overview of the applications of multisets.
\newblock \emph{Novi Sad Journal of Mathematics}, 37\penalty0 (2):\penalty0
  73--92, 2007.
\newblock URL \url{http://eudml.org/doc/226431}.

\bibitem[Stanley(2011)]{stanley_2011}
R.~P. Stanley.
\newblock \emph{Enumerative Combinatorics}, volume~1 of \emph{Cambridge Studies
  in Advanced Mathematics}.
\newblock Cambridge University Press, 2 edition, 2011.
\newblock \doi{10.1017/CBO9781139058520}.

\bibitem[Stanley and Fomin(1999)]{stanley_fomin_1999}
R.~P. Stanley and S.~Fomin.
\newblock \emph{Enumerative Combinatorics}, volume~2 of \emph{Cambridge Studies
  in Advanced Mathematics}.
\newblock Cambridge University Press, 1999.
\newblock \doi{10.1017/CBO9780511609589}.

\bibitem[Van~de Woestijne(2012)]{Woestijne}
C.~E. Van~de Woestijne.
\newblock Factors of disconnected graphs and polynomials with nonnegative
  integer coefficients.
\newblock \emph{Ars Mathematica Contemporanea}, 5\penalty0 (2):\penalty0
  307--323, Apr 2012.
\newblock ISSN 1855-3966.
\newblock \doi{10.26493/1855-3974.202.37f}.
\newblock URL \url{http://dx.doi.org/10.26493/1855-3974.202.37f}.

\bibitem[Yao(1978)]{yao78}
A.~C.-C. Yao.
\newblock New algorithms in bin packing.
\newblock Technical Report CS-TR-1978-662, Stanford University, Department of
  Computer Science, September 1978.

\end{thebibliography}
\label{sec:biblio}

\end{document}